\documentclass[conference]{IEEEtran}
\IEEEoverridecommandlockouts
\usepackage[T1]{fontenc}
\usepackage[utf8]{inputenc}
\usepackage{verbatim}
\usepackage{amsmath}
\usepackage{graphicx}
\usepackage{url}
\usepackage{ifthen}
\usepackage{cite}

\usepackage{amsfonts}
\usepackage[left=1.62cm,right=1.62cm,top=0.75in,bottom=1.1in]{geometry}

\def\BibTeX{{\rm B\kern-.05em{\sc i\kern-.025em b}\kern-.08em
    T\kern-.1667em\lower.7ex\hbox{E}\kern-.125emX}}

\begin{document}
\title{Generalized Multivariate Polynomial Codes for Distributed Matrix-Matrix
Multiplication \thanks{
%This work was partially supported by Grant 2021 SGR 00772 funded by the Universities and Research Department from Generalitat de Catalunya, by the Spanish ministry of economic affairs and digital transformation and the European union - NextGenerationEU UNICO-5G I+D/SUCCESS-6G-VERIFY TSI-063000-2021-41), and by the Spanish Government through the project 6G-AINA, PID2021-128373OB-I00 funded by MCIN/AEI/10.13039/501100011033). This work also received funding from the UK Research and Innovation (UKRI) for the projects AI-R (ERC Consolidator Grant, EP/X030806/1) and INFORMED-AI (EP/Y028732/1).
This work was partially supported by Grant 2021 SGR 00772 funded by Generalitat de Catalunya, by the Spanish Government and the European union through grant - NextGenerationEU UNICO-5G I+D/SUCCESS-6G-VERIFY TSI-063000-2021-41, and by the Spanish Government through the project 6G-AINA, PID2021-128373OB-I00 funded by MCIN/AEI/10.13039/501100011033). This work also received funding from the UK Research and Innovation (UKRI) for the projects AI-R (ERC Consolidator Grant, EP/X030806/1) and INFORMED-AI (EP/Y028732/1).\\
For the purpose of open access, the authors have applied a Creative Commons Attribution (CCBY) license to any Author Accepted Manuscript version arising from this submission.
}
}
\author{\IEEEauthorblockN{Jesús Gómez-Vilardebó\IEEEauthorrefmark{1}, Burak Hasırcıoğlu\IEEEauthorrefmark{2},
and Deniz Gündüz\IEEEauthorrefmark{3}} \IEEEauthorblockA{\IEEEauthorrefmark{1}Centre Tecnològic de Telecomunicacions de Catalunya
(CTTC/CERCA), Barcelona, Spain, jesus.gomez@cttc.es} \IEEEauthorblockA{\IEEEauthorrefmark{2}Alan Turing Institute, London, UK, b.hasircioglu18@imperial.ac.uk}
\IEEEauthorblockA{\IEEEauthorrefmark{3}Imperial College London, London, UK, d.gunduz@imperial.ac.uk}}
\maketitle
\begin{abstract}
Supporting multiple partial computations efficiently at each of the
workers is a keystone in distributed coded computing in order to speed
up computations and to fully exploit the resources of heterogeneous
workers in terms of communication, storage, or computation capabilities.
Multivariate polynomial coding schemes have recently been shown to
deliver faster results for distributed matrix-matrix multiplication
compared to conventional univariate polynomial coding schemes by supporting
multiple partial coded computations at each worker at reduced communication
costs. In this work, we extend multivariate coding schemes to also
support arbitrary matrix partitions. Generalized matrix partitions
have been proved useful to trade-off between computation speed and
communication costs in distributed (univariate) coded computing. We
first formulate the computation latency-communication trade-off in
terms of the computation complexity and communication overheads required
by coded computing approaches as compared to a single server uncoded
computing system. Then, we propose two novel multivariate coded computing
schemes supporting arbitrary matrix partitions. The proposed schemes
are shown to improve the studied trade-off as compared to univariate
schemes. 
\end{abstract}

\section{Introduction}

Matrix-matrix multiplication is a fundamental operation that arises
in many problems, particularly in emerging machine learning applications.
When the input matrices are from large datasets%, as it is increasingly common, 
computing matrix multiplications on a single server within
a reasonable amount of time becomes unfeasible. It is thus essential
to split the job into multiple subtasks and execute them on multiple
servers in parallel. However, servers in modern large distributed
computation clusters usually comprise small, low-end, and unreliable
computational nodes which are severely affected by \textit{``system
noise}'', i.e., faulty behaviors due to computation or memory bottlenecks,
load imbalance, resource contention, hardware issues, etc \cite{Dean2013}.
As a result, task completion times of individual workers become largely
unpredictable, and the slowest workers dominate the overall computation
time. This is referred to in the literature as the \textit{straggler
problem}. %The challenge is thus to secure fast and reliable computations in the face of computation time uncertainty, with reasonable communication and/or storage overheads.

In this work, we address this problem by adopting the \emph{coded
computing} framework initiated in \cite{lee2017speeding,lee2017high,yu2017polynomial,yu2020straggler}.
%For the particular problem of matrix-matrix multiplication, 
A formulation
based on maximum distance separable (MDS) codes was first presented
in \cite{lee2017speeding}. In coded computing, unlike conventional
schemes that rely on subtask repetition, any delayed subtask can be
replaced by any other subtask. As a result, coded-computing-based
solutions provide order-wise improvements
%in terms of 
in task completion
times. In particular, it was possible to recover the original matrix-matrix
product from a number of computations, referred to as \emph{recovery
threshold} that did not increase with the number of workers, but rather
with the total number of matrix block products. \textit{Univariate
polynomial codes}, a form of MDS codes with a particularly efficient
decoding procedure via polynomial interpolation, were described, simultaneously,
in \cite{yu2020straggler} referred to as entangled polynomial codes,
and in \cite{dutta2019optimal} referred to as generalized PolyDot
codes. These schemes offered flexibility to trade the computation
complexity for download communications resources.
%which are used to communicate the results from the workers to the master server. 
Follow-up
works extended the results to secure matrix-matrix multiplications
\cite{Ravi18,Gasp20}, matrix chain multiplications\cite{Fan2021MatrixChain}, among others.
%and coded convolutions \cite{yu2020straggler}. %, among others.

All these works consider a distributed computation model in which
a single subtask is assigned to each worker. To further speed up computations
and to better support heterogeneous workers, subsequent works \cite{amiri2019computation,SongISIT23PartialStragglers,kiani2018exploitation,ozfatura2020straggler,reisizadeh2019coded}
consider instead a distributed computation model with multiple subtasks
per worker. By assigning multiple, smaller subtasks to workers it
can be shown that % left work is left unfinished at workers by the time the computation task has finished,
any partial work conducted by straggler workers can also be exploited.
In \cite{hasirciouglu2022JSAC,hasirciouglu2021JSAC} it was shown
that univariate polynomial coding schemes are inefficient in terms
of the upload communication costs, i.e., the resources needed to communicate
the coded matrices from the master server to workers. To address this
inefficiency, bivariate polynomial codes are proposed in \cite{hasirciouglu2021JSAC}.
These schemes, however, cannot support generalized partition schemes,
failing to trade computation complexity with the download communication
cost.

In this work, we provide an extension of the bivariate codes proposed
in \cite{hasirciouglu2021JSAC} to accommodate generalized matrix
partitions. We introduce two novel multivariate schemes
achieving new points in the trade-off curves between the computation
complexity and upload/download communication overheads.

The remainder of the paper is organized as follows. In Section \ref{sec:system},
the problem formulation is presented. Section
\ref{sec:uni} characterizes the univariate polynomial coding scheme. Next, Section \ref{sec:multi}
introduces the proposed multivariate polynomial codes. Numerical results
are provided in Section \ref{sec:Numerical-Results}. Finally, conclusions
are drawn in Section \ref{sec:conclusions}.

\section{System Model and Problem Formulation}

\label{sec:system} 
\begin{figure}
\begin{centering}
\includegraphics[scale=0.11]{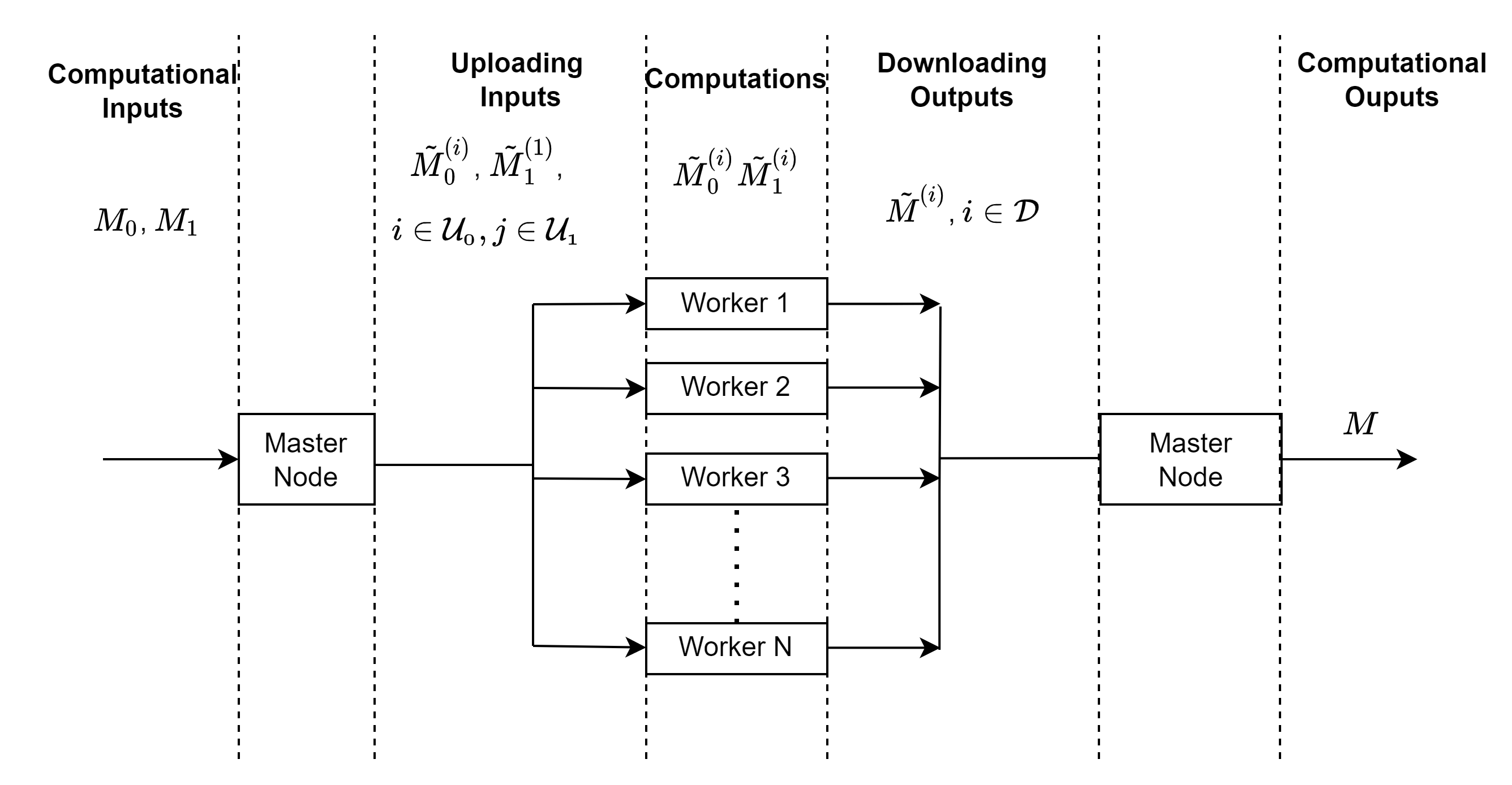}
\par\end{centering}
\caption{Distributed computational system. \label{fig:Computational-system}}
\vspace{-0.5cm}
\end{figure}

We consider the problem of outsourcing the task of multiplying two
large matrices $M_{0}\in\mathbb{F}^{r_{0}\times r_{1}}$ and $M_{1}\in\mathbb{F}^{r_{1}\times r_{2}}$
from a centralized master node to a more computational powerful entity
with $N$ parallel computation nodes/workers, see Figure \ref{fig:Computational-system}.
For that, the master node splits the task into several subtasks and
sends (uploads) appropriate, eg. coded subtask inputs to the nodes.
These nodes in parallel and under a timely coordination from the master,
compute distinct subtasks, and send the results back (downloaded)
to the master node. Multiple subtasks can be assigned to each worker,
in which case, these subtasks are executed sequentially. By exploiting
task partitions and distributing the computation between workers,
the overall computation time can be reduced. However, as we will show
in this work, there is a computation complexity and communication
overhead associated to task partitioning. %
\begin{comment}
We are interested in characterizing the fundamental trade-off between
the upload/download communication costs and the computation latency
of different distributed computation strategies. 
\end{comment}
{} For the uplink communication from the master to the workers, we assume
that each subtask input can be broadcasted to all the workers at the
cost of transmitting only one subtask, i.e. wireless broadcast model.
Workers coordinated by the master, store or discard the subtask inputs
received. %
\begin{comment}
In addition, we assume that only computation inputs that are used
by at least one worker need to be transmitted and thus consume communication
resources. 
\end{comment}
For the downlink, each computation output makes use of separate resources.
\begin{comment}
Improvements might be obtained, for instance, for shared wireless
channels if a way to perform decoding operations exploiting the signal
superposition property of wireless channels is devised. 
\end{comment}
{} %
\begin{comment}
(For clarity, we will dicuss the additional overhead that might be
added if some of these ideas conditions are not met in practice) 
\end{comment}

Specifically, in order to distribute the total computation work, matrix
$M_{0}$ is partitioned into $p_{0}$ partitions vertically and $p_{1}$
partitions horizontally, resulting in $p_{0}p_{1}$ \textit{matrix
block partitions} $M_{0}^{b_{0},b_{1}}\in\mathbb{F}^{\frac{r_{0}}{p_{0}}\times\frac{r_{1}}{p_{1}}}$,
for $b_{0}\in\left[p_{0}\right]$ and $b_{1}\in\left[p_{1}\right]$
where we denoted $\left[n\right]=\left\{ 0,1,\dots,n-1\right\} $.
Similarly, matrix $M_{1}$ is partitioned into $p_{1}$ partitions
vertically, and $p_{2}$ partitions horizontally, resulting in $p_{1}p_{2}$
matrix block partitions $M_{1}^{b_{1},b_{2}}\in\mathbb{F}^{\frac{r_{1}}{p_{1}}\times\frac{r_{2}}{p_{2}}}$
for $b_{1}\in\left[p_{1}\right]$ and $b_{2}\in\left[p_{2}\right]$,
i.e.,

\begin{eqnarray*}
M_{i} & = & \left[\begin{array}{ccc}
M_{i}^{0,0} & \cdots & M_{i}^{0,p_{i+1}-1}\\
\vdots & \ddots & \vdots\\
M_{i}^{p_{i}-1,0} & \cdots & M_{i}^{p_{i}-1,p_{i+1}-1}
\end{array}\right]
\end{eqnarray*}
for $i\in\left\{ 0,1\right\} $. The result of the multiplication
of $M_{0}$ and $M_{1}$, as a function of its matrix blocks, can
be written as

\begin{eqnarray}
M=M_{0}M_{1} & = & \left[\begin{array}{ccc}
M^{0,0} & \cdots & M^{0,p_{2}-1}\\
\vdots & \ddots & \vdots\\
M^{p_{0}-1,0} & \cdots & M^{p_{0}-1,p_{2}-1}
\end{array}\right]\label{eq:matrix result}
\end{eqnarray}
with $M^{n_{0},n_{2}}=\sum_{n_{1}=0}^{p_{1}-1}M^{n_{0},n_{1}}M_{1}^{n_{1},n_{2}}$.
From (\ref{eq:matrix result}) we can readily see that $M$ is comprised
of $p_{0}p_{2}$ matrix block partitions. Given that each matrix block
partition involves $p_{1}$ matrix-matrix multiplications, a total
of $K=p_{0}p_{1}p_{2}$ different block matrix-matrix multiplications
$M_{0}^{n_{0},n_{1}}M_{1}^{n_{1},n_{2}}$ $\left(n_{0},n_{1},n_{2}\right)\in\left[p_{0}\right]\times\left[p_{1}\right]\times\left[p_{2}\right]$
are needed to recover the original multiplication. We refer to the
triplet $\left(p_{0},p_{1},p_{2}\right)$ as the \textit{partition
scheme} and $K$\textit{ as the partition level.}

Given these matrix partitions, an uncoded distributed computation
scheme consists of distributing each of these block matrix-matrix
multiplications into the different parallel workers.
However,
%without any additional measures,
any delay in one of these
computations, results in a delay in the overall computation. This
is known as the \textit{straggler problem} in distributed computing.
A straightforward solution, although inefficient, is \textit{work
repetition}, i.e., replicating each of the block matrix-matrix products
into different workers. Since every replicated computation can only
replace one specific block multiplication, the number of workers needed
grows linearly with the partition level. \textit{Coded computing}
offers a much more efficient approach to this problem. At the master
server, block matrices from $M_{0}$ are linearly encoded into $\left|\mathcal{U}_{0}\right|$
coded block matrices, i.e., $\tilde{M}_{0}^{(i)}\in\mathbb{F}^{\frac{r_{0}}{p_{0}}\times\frac{r_{1}}{p_{1}}}$,
$i\in\mathcal{U}_{0}$, and, similarly, block matrices from $M_{1}$
are linearly encoded into $\left|\mathcal{U}_{1}\right|$ coded block
matrices, i.e., $\tilde{M}_{1}^{(i)}\in\mathbb{F}^{\frac{r_{1}}{p_{1}}\times\frac{r_{2}}{p_{2}}},$$i\in\left|\mathcal{U}_{1}\right|$.
Then, coded block matrices from $M_{0}$ and $M_{1}$ are distributed
to the $N$ workers in the system, which then compute the coded block
matrix-matrix multiplications $\tilde{M}^{(i)}=\tilde{M}_{0}^{(i)}\tilde{M}_{1}^{(i)}\in\mathbb{F}^{\frac{r_{0}}{p_{0}}\times\frac{r_{2}}{p_{2}}}$,
each of which is referred to as a subtask. In order to exploit partial
work, we also consider that each worker may compute multiple subtasks
in a serial manner. That is, as soon as a worker finishes a subtask,
it returns the result to the master for decoding and proceeds to compute
another subtask. After receiving a number of computations known as
the \textit{recovery threshold}, $R_{th}$, the master is able to
decode the target matrix-matrix multiplication, $M_{0}M_{1}$. In
contrast to work repetition, in coded computing, any computation can
be replaced by another that could have been delayed.

\begin{table*}
\caption{Computation complexity and upload/download communications overheads
for different coded distributed computing methods}

\begin{centering}
\begin{tabular}{|c|c|c|c|c|c|}
\hline 
Method$^{*}$  & $R_{th}$  & $\delta$  & $\delta_{u,0}$  & $\delta_{u,1}$  & $\delta_{d}$\tabularnewline
\hline 
\hline 
epc  & $p_{1}p_{2}p_{0}+p_{1}-1$  & $\frac{p_{1}-1}{p_{0}p_{1}p_{2}}$  & $p_{2}-1+p_{2}\delta$  & $p_{0}-1+p_{0}\delta$  & $p_{1}-1+p_{1}\delta$\tabularnewline
\hline 
Bi0  & $p_{0}p_{2}p_{1}+p_{0}\left(p_{1}-1\right)$  & $\frac{p_{1}-1}{p_{0}p_{1}}$  & $\delta$  & $p_{0}-1+p_{0}\delta$  & $p_{1}-1+p_{1}\delta$\tabularnewline
\hline 
Bi2  & $p_{0}p_{1}p_{2}+p_{2}\left(p_{1}-1\right)$  & $\frac{p_{1}-1}{p_{1}p_{2}}$  & $p_{2}-1+p_{2}\delta$  & $\delta$  & $p_{1}-1+p_{1}\delta$\tabularnewline
\hline 
Tri  & $p_{0}p_{2}p_{1}+p_{0}p_{2}\left(p_{1}-1\right)$  & $\frac{p_{1}-1}{p_{1}}$  & $\delta$  & $\delta$  & $p_{1}-1+p_{1}\delta$\tabularnewline
\hline 
\end{tabular}\\
 {\footnotesize{}{}{$^{*}$See Section III and IV for the definition
of these abreviations.} }{\footnotesize\par}
\par\end{centering}
\begin{centering}
 
\par\end{centering}
{\footnotesize{}{}\label{Table:overheads}
 }{\footnotesize\par}

 \vspace{-0.5cm}
\end{table*}

Although coded computing is an efficient method to %
\begin{comment}
combat the straggler problem and
\end{comment}
achieve lower \textit{computation latencies}, it incurs \textit{communication}
and \textit{computation complexity overheads}. %
\begin{comment}
In the following, we define the trade-off between the average \textit{computation
latency}, also referred to as average \textit{completion time}, the
computation complexity overheads, and the upload/download communication
overheads in coded computing. For that, first, we define each of these
terms.
\end{comment}
{} %
\begin{comment}
together with the main assumptions adopted in our model. 
\end{comment}
The \textit{average computation latency} is here defined as the average
time taken to obtain the computation result. %
\begin{comment}
For coded computing, this includes the time required to encode, upload
the coded matrices to the workers, computing $R_{th}$ block matrix-matrix
products distributively at workers, downloading the results from the
workers to the masters, and decoding at the master. However, 
\end{comment}
For simplicity, we assume the computation latency
is mainly dominated by the computation time at workers.
%, and that all
%other contributions are either negligible or constant. %
\begin{comment}
The average computation latency then depends, mainly, on the recovery
threshold, $R_{th}$, i.e., the total number of computations that
are needed, the number of parallel workers, $N$, and the statistics
of the computation time for each subtask at workers.
\end{comment}
{} %
\begin{comment}
, for which its first order statistics depend on the computation complexity 
\end{comment}
We get lower computation latency as we increase the number of parallel
workers, $N$, since there is more work parallelization, but also
as we increase the partition level, $K$, since then each subtask
becomes smaller and less work is left unfinished at the workers by
the time $R_{th}$ computations are completed.

For the overheads, we will use as a reference, the setting where we
outsource the computation to a single server, which we refer to as
SS. Then, we define the \textit{computation complexity overhead},
$\delta$, as the increment in computation complexity, i.e., number
of element-wise multiplications, required by coded computing $C^{\text{CC}}$,
relative to the computation complexity in a single server, $C^{\text{SS}}$.
That is, $C^{\text{CC}}=C^{\text{SS}}\left(1+\delta\right)$. For
the uncoded single server setting, multiplication involves, roughly,
$C^{\text{SS}}=r_{0}r_{1}r_{2}$ element-wise multiplications
%\footnote{More precisely, the commonly used cubic algorithm achieves a complexity of $\mathcal{O}\left(r_{0}r_{1}r_{2}\right)$ for the general case. Reduced complexity is possible under particular conditions, however, all known approaches requires a super quadratic complexity.}
. Likewise, for coded computing, each partial computation at the workers
involves $\frac{r_{0}}{p_{0}}\frac{r_{1}}{p_{1}}\frac{r_{2}}{p_{1}}$
element-wise multiplications. Given that $R_{th}$ subtasks are required,
the computation complexity is $C^{\text{CC}}=R_{th}\frac{r_{0}}{p_{0}}\frac{r_{1}}{p_{1}}\frac{r_{2}}{p_{1}}$,
and thus the computation overheads is 
\begin{eqnarray}
\delta & = & \frac{C^{\text{CC}}}{C^{\text{SS}}}-1=\frac{R_{th}}{K}-1.\label{eq:overhead_complexity}
\end{eqnarray}
\begin{comment}
Consequently, if $R_{th}>K$, then coded computing requires overall
more computation work than the original matrix-matrix product on a
single server. For a computation system with a coded computation overhead
of $\delta$, we say that coded computation requires $(100\cdot\delta)\%$
of additional multiplications compared to the uncoded computation
in a single server.
\end{comment}

The \textit{download communication overhead}, $\delta_{d}$, is similarly
defined as the increment of the cost of communicating the block matrix
products from the workers to the master in coded computing, $C_{d}^{\text{CC}}$,
relative to the cost of communicating the original product result
from a single server, $C_{d}^{\text{SS}}$. That is $C_{d}^{\text{CC}}=C_{d}^{\text{SS}}\left(1+\delta_{d}\right)$.
If we measure the communication cost in units of the size of the entries
of the resultant product matrix, we have that, since $R_{th}$ block
matrix products, i.e., subtasks, are returned, each of size $\frac{r_{0}}{p_{0}}\frac{r_{2}}{p_{2}}$,
the cost is $C_{d}^{\text{CC}}=R_{th}\frac{r_{0}r_{2}}{p_{0}p_{2}}$.
On the other hand, given that the size of the matrix product result
is $r_{0}r_{2},$ the cost of communicating the result from a single
server is $C_{d}^{\text{SS}}=r_{0}r_{2}$, and thus, 
\begin{eqnarray}
\delta_{d} & = & \frac{R_{th}}{p_{0}p_{2}}-1=\left(p_{1}-1\right)+p_{1}\delta.\label{eq:downcost2}
\end{eqnarray}
In (\ref{eq:downcost2}), we can distinguish two different contributions,
one related with computation complexity overhead, $\delta$, and another
related with the partition level \textit{$p_{1}$.}%
\begin{comment}
{[}PROVIDE FURTHER DISCUSSION ON WHY THE CONTRIBUTION OF p1 is fundamental
here-{]} 
\end{comment}
{} 

The \textit{upload communication overhead} is similarly defined as
the increment in the cost of communicating the coded block matrices
from the master to the workers, relative to the cost of communicating
the original matrices $M_{0}$ and $M_{1}$ to a single server. %
\begin{comment}
The exact upload communication cost depends on the specific computation
scheduling model. In this work, we adopt an online scheduling model
in which the coded matrix partitions are broadcasted to the workers
as soon as they are needed by one or more of the workers.
\end{comment}
When the computation has finished, we have obtained $R_{th}$ block
matrix products and $N-1$ are left unfinished at workers. Thus, we
need to upload the necessary inputs to run, at least, $R_{th}+N-1$
computations. For simplicity, we assume the number of partitions is
large, and it is satisfied that $R_{th}\gg N$, in that case, we can
closely approximate %\footnote{This cost would be a good approximation for highly coordinated distributed computing systems where every worker frequently monitors the progressof computations at every worker, and only requests new coded matrix when needed.}
\begin{comment}
\end{comment}
the upload communication overheads by counting the number of coded
matrix block $\tilde{M}_{0}$ and $\tilde{M}_{1}$ denoted as $R_{0}$
and $R_{1}$, respectively, that need to be sent to the workers in
order to obtain exactly $R_{th}$, block matrix products.%
\begin{comment}
For the sake of simplicity, we are thus ignoring all the upload communication
that are only used by the computations left unfinished at workers,
after the first $R_{th}$ block products are returned to the master. 
\end{comment} Given that the coded blocks $\tilde{M}_{0}$ and $\tilde{M}_{1}$
have size $\frac{r_{0}}{p_{0}}\frac{r_{1}}{p_{1}}$ and $\frac{r_{1}}{p_{1}}\frac{r_{2}}{p_{2}}$,
respectively we have that the uploading cost associated with sending
the $R_{0}$ code blocks from $M_{0}$ and $R_{1}$ code blocks from
$M_{1}$ are $C_{u,0}^{\text{CC}}=R_{0}\frac{r_{0}r_{1}}{p_{0}p_{1}}$
and $C_{u,1}^{\text{CC}}=R_{1}\frac{r_{1}r_{2}}{p_{1}p_{2}}$, respectively.
On the other hand, given that original matrices have size $r_{0}r_{1}$
and $r_{1}r_{2},$ the cost of uploading them to a single server is
$C_{U,0}^{\text{SS}}=r_{0}r_{1}$ and $C_{U,1}^{\text{SS}}=r_{1}r_{2}$,
respectively. The upload communication overheads are then obtained
from $C_{U,0}^{\text{CC}}=C_{U,0}^{\text{SS}}\left(1+\delta_{u,0}\right)$,
and $C_{U,1}^{\text{CC}}=C_{U,1}^{\text{SS}}\left(1+\delta_{u,1}\right)$
as 
\begin{eqnarray}
\delta_{u,0} & = & \frac{R_{0}}{p_{0}p_{1}}-1=p_{2}\frac{R_{0}}{R_{th}}\left(\delta+1\right)-1\label{eq:up0oh2}\\
\delta_{u,1} & = & \frac{R_{1}}{p_{1}p_{2}}-1=p_{0}\frac{R_{0}}{R_{th}}\left(\delta+1\right)-1\label{eq:up1oh2}
\end{eqnarray}
where (\ref{eq:up0oh2}) and (\ref{eq:up1oh2}) follows from (\ref{eq:overhead_complexity}).

As a reference for the rest of the paper, in Table \ref{Table:overheads}
we summarize the obtained overheads %recovery thresholds, computation, and upload/download
%communications overheads obtained 
for the different coded computing strategies discussed in this work.

\begin{comment}
WE CONSIDER THE MINUMUM NUMBER OF UPLADED MATRICES

If we take into account the uploaded cost associated to unfinished
computaitn we have mainl N-1 compuation. This is does not apper in
the signle serve case 
\end{comment}

\section{Univariate Polynomial Codes}

\label{sec:uni}

As a reference, we first provide the computation complexity and communication
overheads for a more general version of the univariate polynomial
coding schemes considered in the literature, e.g., entangled polynomial
codes presented in \cite{dutta2019optimal,dutta2018unified} and \cite{yu2020straggler}.
The original coding scheme was described assuming that only one coded
block matrix product is assigned to each worker. Here, we describe
its direct extension to support multiple coded computations per worker
that are executed sequentially. This extension provides more flexibility
to the system design, as now the number of matrix partitions is not
limited by the number of parallel workers in the system.

With entangled polynomial codes (epc), the block matrices $M_{0}^{(b_{0},b_{1})}$
and $M_{1}^{(b_{1},b_{2})}$ are encoded with the polynomials 
\begin{eqnarray*}
\tilde{M}_{0}^{\text{epc}}(x) & = & \sum_{b_{0}=0}^{p_{0}-1}\sum_{b_{1}=0}^{p_{1}-1}M_{0}^{(b_{0},b_{1})}x^{p_{1}p_{2}b_{0}+b_{1}}\\
\tilde{M}_{1}^{\text{epc}}(x) & = & \sum_{b_{2}=0}^{p_{2}-1}\sum_{b_{1}=0}^{p_{1}-1}M_{1}^{(p_{1}-1-b_{1},b_{2})}x^{p_{1}b_{2}+b_{1}}.
\end{eqnarray*}
Observe that within the coefficients of the univariate product polynomial
$\tilde{M}^{\text{epc}}\left(x\right)=\tilde{M}_{0}^{\text{epc}}(x)\tilde{M}_{1}^{\text{epc}}(x)$,
we can find the matrix blocks of the target matrix product. In particular,
$M^{n_{0},n_{2}}=\sum_{n_{1}=0}^{p_{1}-1}M_{0}^{n_{0},n_{1}}M_{1}^{n_{1},n_{2}}$
is given by the coefficient of the monomial $x^{p_{1}p_{2}n_{0}+p_{1}n_{2}+p_{1}-1}$
of $\tilde{M}^{\text{epc}}\left(x\right)$. Moreover, the product
polynomial has degree $p_{1}p_{2}p_{0}+p_{1}-2$ and thus, via polynomial
interpolation we can recover its coefficients from $R_{th}^{\text{epc}}=p_{1}p_{2}p_{0}+p_{1}-1$
evaluations of $M\left(x\right)$. To obtain these evaluations in
a distributed manner, at least $R_{th}$ pairs of evaluations of the
coded matrix blocks $\left(\tilde{M}_{0}^{\text{epc}}(x_{i}),\tilde{M}_{1}^{\text{epc}}(x_{i})\right)$
for $x_{1}=1,\dots,R_{th}$ need to be uploaded to the workers, which
then compute the products $\tilde{M}^{\text{epc}}\left(x_{i}\right)=\tilde{M}_{0}^{\text{epc}}(x_{i})\tilde{M}_{1}^{\text{epc}}(x_{i})$
and send them back to the master for decoding. The computation complexity
overhead can be obtained from (\ref{eq:overhead_complexity}) as $\delta^{\text{epc}}=\frac{p_{1}-1}{p_{1}p_{2}p_{0}}.$
The upload communication overhead is obtained by particularizing (\ref{eq:up0oh2})
and (\ref{eq:up1oh2}) with $R_{0}^{\text{epc}}=R_{1}^{\text{epc}}=R_{th}^{\text{epc}},$
as $\delta_{u,0}^{\text{epc}}=p_{2}-1+p_{2}\delta^{\text{epc}}$ and
$\delta_{u,1}^{\text{epc}}=p_{0}-1+p_{0}\delta^{\text{epc}}$. The
download communication overhead is directly given by (\ref{eq:downcost2}).

\begin{comment}
If there were no computation complexity overheads and no constraints
on the communication overheads, then the average computation latency
would be minimized by choosing the highest possible subtask partition
level $K=p_{0}p_{1}p_{2}$. However, 
\end{comment}
Observe that zero computation complexity overhead is achieved only
when $p_{1}=1$. Alternatively, when we have $p_{1}>1$, we can achieve
a negligible complexity overhead by increasing $p_{0}$ and $p_{2}$,
at the cost of increasing the download and upload communication overheads.
In the following, we describe two multivariate coded computing schemes,
which support arbitrary matrix partitioning, and remove the penalization
from $p_{0}$ and $p_{2}$ on the upload communication overheads.

\begin{comment}
{[}Include critic discussion on the tradeoff presented in \cite{dutta2019optimal}.
There this scheme was motivated as mean to trade increasing download
communication cost (aka overheads) to reduce the recovery threshold
(aka complexity overhead), ($R_{th},$$C_{d}^{\text{epc}}$), vs $\left(\delta^{\text{epc}},\delta_{d}^{\text{epc}}\right)$.{]} 
\end{comment}

\vspace{-0.1cm}
\section{Multivariate Coded Computing}
\label{sec:multi}

In \cite{hasirciouglu2021JSAC}, a bivariate polynomial coding scheme
was presented supporting partial computation at workers and improving
the upload communication costs, as compared to univariate coding schemes
for the case $p_{1}=1$. Here, we extend this scheme to support the
case $p_{1}\geq1.$ %We introduce two novel multivariate polynomial codes one based on bivariate and the other on tri-variate polynomial codes\textcolor{red}{.}
In contrast to univariate polynomial codes of degree $d$, for which,
any set of $d+1$ distinct evaluation points guarantee decodability,
for multi-variate polynomials, there exists only a few known sets
of multi-variate evaluation points for which decodability can be guaranteed.
With an independent random choice of the evaluation points we can
achieve almost decodability, that is, decodability with probability
increasing with the size of the operation field. The interested reader
is referred to \cite{lorentz2006multivariate} for details on multivariate
polynomial interpolation and to \cite{hasirciouglu2021JSAC,hasirciouglu2022JSAC}
for its application to coded computing. However, as we will see here,
to achieve low upload communications overheads, the evaluation set
must have structure. For simplicity, in this work we restrict the
analysis to the multivariate Cartesian product evaluation set. The
Cartesian product set guarantees decodability and provides the lowest
possible upload communication overheads. However, it may require large
storage capacity at the workers in offline upload communication settings,
i.e. if all the subtask inputs are uploaded before any computation
starts at workers. 
%The study of alternative evaluation sets with better trade-off between storage and upload communication overheads is leftfor future work. 

\subsection{Bivariate Polynomial Codes}

First, we present bivariate coding polynomials, referred to as Bi0,
where the coded matrix partitions associated to $M_{0}$ are encoded
with a bivariate polynomial and the ones associated to $M_{1}$ with
a univariate polynomial given by 
\begin{eqnarray*}
\tilde{M}_{0}^{\text{Bi0}}\left(x,y\right) & = & \sum_{b_{0}=0}^{p_{0}-1}\sum_{b_{1}=0}^{p_{1}-1}M_{0}^{\left(b_{0},b_{1}\right)}x^{b_{0}}y^{p_{1}-1-b_{1}}\\
\tilde{M}_{1}^{\text{Bi0}}\left(y\right) & = & \sum_{b_{2}=0}^{p_{2}-1}\sum_{b_{1}=0}^{p_{1}-1}M_{1}^{(b_{1},b_{2})}y^{b_{2}p_{1}+b_{1}}.
\end{eqnarray*}
Observe that the bivariate product polynomial $\tilde{M}^{\text{Bi0}}(x,y)=\tilde{M}_{0}^{\text{Bi0}}\left(x,y\right)\tilde{M}_{1}^{\text{Bi0}}\left(y\right)$
contains the matrix blocks $M^{n_{0},n_{2}}$ of the product matrix
$M$ as the coefficients of the monomials $x^{n_{0}}y^{p_{1}-1+n_{2}p_{1}}$.
The product polynomial has a degree $p_{0}-1$ on $x$ and $p_{2}p_{1}+p_{1}-2$
on $y$, and thus, via bivariate polynomial interpolation we can recover
the coefficients of $\tilde{M}^{\text{Bi0}}(x,y)$ from $R_{th}^{\text{Bi0}}=p_{0}\left(p_{2}p_{1}+p_{1}-1\right)$
evaluations. Specifically, consider the bi-variate Cartesian product
evaluation set, that is $\left(x,y\right)\in\mathcal{X}\times\mathcal{Y}$,
with $\left|\mathcal{X}\right|=p_{0},$ and $\left|\mathcal{Y}\right|=p_{2}p_{1}+p_{1}-1$.
For this choice of evaluation points, the server broadcasts $\tilde{M}_{1}^{\text{Bi0}}(y)$
for each $y\in\mathcal{\mathcal{Y}}$ and $\tilde{M}_{0}^{\text{Bi0}}(x,y)$
for each $\left(x,y\right)\in\mathcal{X}\times\mathcal{Y}$, and thus
$R_{0}^{\text{Bi0}}=\left(p_{2}p_{1}+p_{1}-1\right)p_{0}=R_{th}^{\text{Tri}}$,
and $R_{1}^{\text{Bi0}}=p_{2}p_{1}+p_{1}-1=\frac{R_{th}^{\text{Tri}}}{p_{0}}$.
Then, the workers, timely coordinated by the master, compute all the
products $\tilde{M}^{\text{Bi0}}\left(x,y\right)$ with $\left(x,y\right)\in\mathcal{X}\times\mathcal{Y}$.
Thus, the upload communication overheads from \eqref{eq:up0oh2} and
\eqref{eq:up1oh2} are given by %\begin{eqnarray*}
%\delta_{u,0}^{\text{Bi0}} & = & p_{2}-1+p_{2}\delta^{\text{Bi0}}\\
%\delta_{u,1}^{\text{Bi0}} & = & \delta^{\text{Bi0}}
%\end{eqnarray*}
$\delta_{u,0}^{\text{Bi0}}=p_{2}-1+p_{2}\delta^{\text{Bi0}}$ and
$\delta_{u,1}^{\text{Bi0}}=\delta^{\text{Bi0}}$, where the computation
complexity overhead from (\ref{eq:overhead_complexity}) is given
by $\delta^{\text{Bi}0}=\frac{p_{1}-1}{p_{1}p_{2}}.$ Compared to
the univariate polynomial coding, the upload communications overheads
are reduced because now each evaluation of $\tilde{M}_{1}^{\text{Bi0}}(y)$
can be multiplied with up to $p_{0}$ evaluations of $\tilde{M}_{0}^{\text{Bi0}}(x,y)$,
with a common $y-$coordinate. Finally, the download communication
overhead is directly given by (\ref{eq:downcost2}) as shown in Table
\ref{Table:overheads}. 

Notice that in an offline upload communication setting we must have sufficient storage capacity to compute any of
the subtasks during all computation time. The storage requirements
could be reduced in practice with online settings where new evaluations
are sent as computation advances, replacing previous evaluations that
are not needed for future computations at a particular worker. By interchanging the encoding polynomials for $M_{0}$ and $M_{1}$,
we can define a dual bivariate scheme referred to as Bi2 with the
recovery thresholds and overhead expressions that result from interchanging
$p_{0}$ and $p_{2}$ as shown in Table \ref{Table:overheads}.
\begin{comment}
The choice between Bi0 or Bi2 to minimize the computation latency
under fixed communication overhead constraints in a particular setting
is not straightforward. Note that if $p_{2}>p_{0}$, selecting Bi0
over Bi2 would result in lower computation complexity and download
overheads. However, at the cost of increasing the upload overheads.
On the other hand, if $p_{0}>p_{2}$, the opposite holds true.
\end{comment}

\subsection{Tri-variate Polynomial Codes}

Compared to the univariate epc scheme, with bivariate coding schemes,
we were able to reduce the upload communication overheads on one of
the two coded matrices by eliminating the contribution of $p_{0}-1$
in $\delta_{u,1}$ or the contribution of $p_{2}-1$ in $\delta_{u,0}$.
This was, however, at the cost of increasing the computation complexity
overhead by a factor of $p_{0}$ or $p_{2}$, respectively. Next,
we present a tri-variate polynomial coding scheme for which the upload
communication overhead is reduced for both matrices simultaneously
at the expense of increasing the computation complexity overhead by
a factor of $p_{0}p_{2}$ relative to the epc scheme.{} %
\begin{comment}
The tri-variate scheme reduces exactly to the bivariate scheme in
{[}Burak{]} for the case $p_{1}=1$. 
\end{comment} For the tri-variate polynomial coding scheme, the block matrix inputs
are encoded with
\begin{eqnarray*}
\tilde{M}_{0}^{\text{Tri}}(x,y) & = & \sum_{b_{0}=0}^{p_{0}-1}\sum_{b_{1}=0}^{p_{1}-1}M_{0}^{(b_{0},b_{1})}x^{b_{0}}y^{b_{1}}\\
\tilde{M}_{1}^{\text{Tri}}(y,z) & = & \sum_{b_{2}=0}^{p_{2}-1}\sum_{b_{1}=0}^{p_{1}-1}M_{1}^{(p_{1}-1-b_{1},b_{2})}y^{b_{1}}z^{b_{2}}.
\end{eqnarray*}
Observe that within the coefficients of the tri-variate product polynomial
$\tilde{M}^{\text{Tri}}(x,y,z)=\tilde{M}_{0}^{\text{Tri}}(x,y)\tilde{M}_{1}^{\text{Tri}}(y,z)$,
we can find the matrix blocks of the desired matrix product. Specifically,
$M^{n_{0},n_{2}}$ is given by the coefficient of the monomial $x^{n_{0}}y^{p_{1}-1}z^{n_{2}}$
in $\tilde{M}^{\text{Tri}}(x,y,z)$. Moreover, the product polynomial
has degree $p_{0}-1$ in $x$, $p_{2}-1$ in $z$, and $2p_{1}-2$
in $y$. Thus, via multivariate polynomial interpolation it is potentially
possible to recover its coefficients from $R_{th}^{\text{Tri}}=p_{0}p_{2}\left(2p_{1}-1\right)$
evaluations of $\tilde{M}^{\text{Tri}}(x,y,z)$. To obtain these evaluations,
we consider the Cartesian product set, $\mathcal{X}\times\mathcal{Y}\times\mathcal{Z}$,
with $\left|\mathcal{X}\right|=p_{0},$ $\left|\mathcal{Y}\right|=2p_{1}-1$
and $\left|\mathcal{Z}\right|=p_{2}$. Then, the server broadcasts
$\tilde{M}_{0}^{\text{Tri}}(x,y)$ for each $\left(x,y\right)\in\mathcal{X}\times\mathcal{Y}$,
and thus $R_{0}^{\text{Tri}}=\left(2p_{1}-1\right)p_{0}=\frac{R_{th}^{\text{Tri}}}{p_{2}}$,
and $\tilde{M}_{1}^{\text{Tri}}(y,z)$ for each $\left(y,z\right)\in\mathcal{\mathcal{Y}\times\mathcal{Z}}$,
and thus $R_{1}^{\text{Tri}}=\left(2p_{1}-1\right)p_{2}=\frac{R_{th}^{\text{Tri}}}{p_{0}}$.
Then workers, coordinated by the master can compute one-by-one all
the products $\tilde{M}^{\text{Tri}}\left(x,y,z\right)$. Thus, from
(\ref{eq:up0oh2}) and (\ref{eq:up1oh2}), the upload communication
overheads are given by %\begin{eqnarray*}
%\delta_{u,0}^{\text{Tri}} & = & \delta^{\text{Tri}}\\
%\delta_{u,1}^{\text{Tri}} & = & \delta^{\text{Tri}}.
%\end{eqnarray*}
$\delta_{u,0}^{\text{Tri}}=\delta^{\text{Tri}}$ and $\delta_{u,1}^{\text{Tri}}=\delta^{\text{Tri}}$.
\begin{comment}
Here, for simplicity, we are again assuming an offline setting, where
workers have sufficient storage capacity to compute any of the subtasks
during all computation time.
\end{comment}
Compared to the univariate polynomial coding, the upload communications
overheads are reduced because now each evaluation of $\tilde{M}_{0}^{\text{Tri}}(x,y)$
can be multiplied with up to $p_{2}$ evaluations of $\tilde{M}_{1}^{\text{Tri}}(y,z)$,
with a common $y-$coordinate. Similarly, each evaluation of $\tilde{M}_{1}^{\text{Tri}}(y,z)$
can be multiplied with up to $p_{0}$ evaluations of $\tilde{M}_{0}^{\text{Tri}}(x,y)$
with a common $y-$coordinate. Finally, the download communication
overhead is again directly given by (\ref{eq:downcost2}) as shown
in Table \ref{Table:overheads}.

This encoding polynomials $\tilde{M}_{0}^{\text{Tri}}(x,y)$ and $\tilde{M}_{1}^{\text{Tri}}(y,z)$
have previously appeared in \cite{dutta2019optimal} to provide a
unified formulation to different uni-variate polynomial coding schemes,
including Generalized PolyDot Codes in \cite{dutta2019optimal} or,
the epc codes \cite{yu2020straggler}. For that, the tri-variate polynomials
were reduced to uni-variate polynomials by choosing evaluation points
on single dimensional curves, see \cite[Table I]{dutta2019optimal}.
Specifically, to recover the epc encoding polynomial, we choose $x=y^{p_{1}p_{2}}$
and $z=y^{p_{1}}$. However, the potential of their multi-variate
evaluation, witch we address in this work, was not considered.

\section{Numerical Results\label{sec:Numerical-Results}}

We are interested in characterizing the trade-off between the average
computation latency and the upload/download communication overhead. For that, we search for the partition scheme $\left(p_{0},p_{1},p_{2}\right)$
that minimizes the average computation latency $T(p_{0},p_{1},p_{2})$
while guaranteeing bounded upload/download communication overheads.
That is, 
\begin{eqnarray}
 & \min_{p_{0},p_{1},p_{2}} & T(p_{0},p_{1},p_{2})\label{eq:problem}
\end{eqnarray}
s.t. $\delta_{u,0}\leq\hat{\delta}_{u,0}$, $\delta_{u,1}^{\text{}}\leq\hat{\delta}_{u,1}$,
and $\delta_{d}\leq\hat{\delta}_{d}$, where $\hat{\delta}_{u,0}$,
$\hat{\delta}_{u,1}$ and $\hat{\delta}_{d}$ are fixed system constraints
on upload and download costs. We solve the problem via Monte Carlo simulations.
To model the subtask completion time at workers, we consider the shifted
exponential model, which is a common model employed in distributed
computing \cite{lee2017speeding,liang2014tofec}. Suppose that the
completion of the full task in a single server has an average completion
time given by $T_{\text{full}}=T_{0}+T_{E}$ where $T_{0}$ is a constant
and $T_{E}$ is exponentially distributed random variable with parameter
$\lambda$. Then, %\[
%\text{F}(t)=\text{Pr}(T_{full}\leq t)=1\text{–\ensuremath{\exp}\ensuremath{\left(-\lambda\left(t-T_{0}\right)\right)}.}
%\]
%We further assume that 
by partitioning the full task into $K$ subtasks, the cumulative distribution
function of the completion time of each individual subtask $T_{i}$
is 
\begin{eqnarray*}
\text{F}_{i}(t) & = & P(T_{i}\leq t)=1\text{–}\ensuremath{\exp}\left(-\lambda K\left(t-\frac{T_{0}}{K}\right)\right)
\end{eqnarray*}
and thus, $T_{i}$ is the sum of a constant $T_{0,i}=\frac{T_{0}}{K}$
plus an exponential random variable with parameter $\lambda_{i}=\lambda K$.

\begin{figure}
\begin{centering}
\includegraphics[scale=0.55]{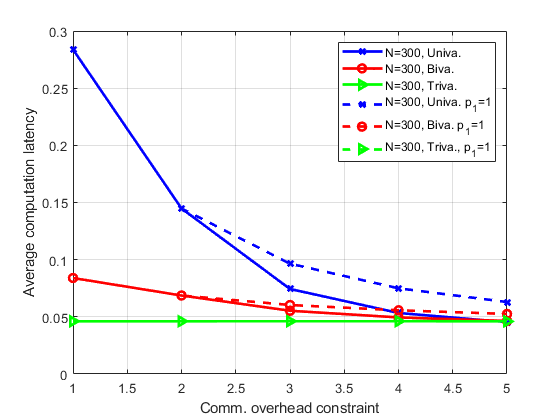} 
\par\end{centering}
\caption{Average computation latency as a function of the communication overhead
constraint for the different coding schemes.\label{fig:Average-completion-time}}
\end{figure}

We search over all partition schemes $\left(p_{0},p_{1},p_{2}\right)$
that satisfy all the communication overhead constraints. Here we consider
the particular case when $\hat{\delta}_{u,0}=\hat{\delta}_{u,1}=\hat{\delta}_{d}=\hat{\delta}_{u,d}$.
To reduce the search space, we further limit the maximum partition
levels to satisfy $p_{0}\leq\hat{p}_{0}$ and $p_{2}\leq\hat{p}_{2}$
, while $p_{1}$ is left unbounded. As an illustrative case, we choose
$\frac{1}{\lambda}=10$ and $T_{0}=\frac{1}{10\lambda}$, $N=300$
workers and $\hat{p}_{0}=\hat{p}_{2}=10$. In Figure \ref{fig:Average-completion-time},
we show the average computation latency as a function of the communication
overhead constraint for the three coded computing schemes. As a reference,
we also show the results obtained by forcing $p_{1}=1$ (dotted lines).
We can observe that the tri-variate scheme obtains the best performance
in this situation, while univariate scheme is heavily penalized at
low communication overheads.

\section{Conclusion}

\label{sec:conclusions}

We formulated the computation latency versus communication/complexity
overheads trade off analysis in distributed coded computing systems.
The overheads are defined relative to the resources required by outsourcing
the computation to a single server in an uncoded manner. We presented
two novel multivariate polynomial coding schemes supporting arbitrary
matrix partitions, which, compared to univariate polynomial codes,
require lower upload communication overheads at the cost of increasing
the computation complexity. Simulation results show significant improvements
in the computation latency for communication overhead constrained
systems.%
\begin{comment}
The reduction of the downloading communication overhead is left often
to further research. 
\end{comment}

%%%%%%
%% Appendix:
%% If needed a single appendix is created by
%%
%\appendix
%%
%% If several appendices are needed, then the command
%%
% \appendices
%%
%% in combination with further \section commands can be used.
%%%%%%

%\bibliographystyle{IEEEtran}
%\bibliography{definitions,bibliofile}
%%
%% where we here have assumed the existence of the files
%% definitions.bib and bibliofile.bib.
%% BibTeX documentation can be obtained at:
%% http://www.ctan.org/tex-archive/biblio/bibtex/contrib/doc/
%%%%%%

%% Or you use manual references (pay attention to consistency and the
%% formatting style!):

 \bibliographystyle{IEEEtran}
\bibliography{IEEEabrv,misRef}

\end{document}